**Noise-plasticity correlations of gene expression in the multicellular organism**

***Arabidopsis thaliana***

Koudai Hirao[1], Atsushi J Nagano[2,3,4] Akinori Awazu[1,5]

[1]Department of Mathematical and Biosciences, Hiroshima University, Kagamiyama 1-3-1, Higashi-Hiroshima, Hiroshima, 739-8526, Japan

[2]Faculty of Agriculture, Ryukoku University, Yokatani 1-5, Seta Ohe-cho, Otsu-shi, Shiga 520-2194, Japan

[3]Center for Ecological Research, Kyoto University, Hirano 509-3-2 Otsu, Shiga, 520-2113, Japan

[4]JST PRESTO, Honcho 4-1-8, Kawaguchi, Saitama, 332-0012, Japan

[5]Research Center for Mathematics on Chromatin Live Dynamics, Hiroshima University, Kagamiyama 1-3-1, Higashi-Hiroshima, Hiroshima, 739-8526, Japan

correlation, multicellular organism, *Arabidopsis thaliana*

**Abstract**

Gene expression levels exhibit stochastic variations among genetically identical organisms under the same environmental conditions (called gene expression "noise" or phenotype "fluctuation"). In yeast and *Escherichia coli*, positive correlations have been found between such gene expression noise and "plasticity" with environmental variations. To determine the universality of such correlations in both unicellular and multicellular organisms, we focused on the relationships between gene expression "noise" and "plasticity" in *Arabidopsis thaliana*, a multicellular model organism. In recent studies on yeast and *E. coli*, only some gene groups with specific properties of promoter architecture, average expression levels, and functions exhibited strong noise-plasticity correlations. However, we found strong noise-plasticity correlations for most gene groups in *Arabidopsis*; additionally, promoter architecture, functional essentiality of genes, and circadian rhythm appeared to have only a weak influence on the correlation strength. The differences in the characteristics of noise-plasticity correlations may result from three-dimensional chromosomal structures and/or circadian rhythm.

**Introduction**

In many organisms, stochastic variations in gene expression have been observed among individuals in a genetically identical population under constant environmental conditions (Elowitz et al., 2002; Furusawa et al., 2005; Golding et al., 2005; Kaern et al., 2005; Newman et al., 2006; Chang et al., 2008; Konishi et al., 2008; Taniguchi et al., 2010; So et al., 2011; Silander et al., 2012; Woods, 2014). These variations, called "noise," differ among genes with some genes being more prone to displaying this type of behavior. Recent investigations in *Escherichia coli* and the budding yeast *Saccharomyces cerevisiae* reported that the magnitude of expression noise shown by a gene can exhibit a positive correlation with "plasticity", the variation in its expression levels due to mutation or environmental change (Sato et al., 2003; Blake et al., 2003; Landry et al., 2007; Choi and Kim, 2008, 2009; Tirosh and Barkai, 2008; Lehner, 2010; Lehner and Kaneko, 2011; Bajic and Poyatos, 2012; Singh, 2013). Such noise-plasticity correlations in gene expression levels reflect both a strategy and a mechanism for regulation of several gene groups. For example, in *E. coli* and yeast, some gene groups with specific properties of promoter architecture, average expression levels, and functions exhibit stronger noise-plasticity correlations (Tirosh and Barkai, 2008; Lehner, 2010; Bajic and Poyatos, 2012; Singh, 2013). Thus, the consideration of such noise-plasticity correlations in genes may provide insights into the survival strategies of various organisms.

The studies mentioned above in budding yeast and *E. coli* identified several aspects of gene expression noise-plasticity correlations that occur in unicellular organisms. In multicellular organisms, gene expression patterns in each cell are modified by intercellular interactions that are mediated through several signaling molecules. Recently, gene expression noise was shown to play important roles in sustaining pluripotency of embryonic stem cells, in cell fate decisions, and in cell differentiation in multicellular organisms (Mitsui et al., 2003; Kaneko, 2006; Kalmar et al., 2009; Ochiai et al., 2014). However, it remains to be elucidated whether gene expression noise and plasticity at an individual level in multicellular organisms depend on similar characteristics to those in unicellular organisms. In this study, as a first step to the elucidation of the gene expression noise-plasticity relationships in multicellular organisms, we analyzed transcriptome data from the multicellular plant model, *Arabidopsis thaliana*, and compared the results to those of typical unicellular organisms, i.e., budding yeast and *E. coli*.

The physiological properties of *Arabidopsis* have been extensively investigated at both the whole plant level and the tissue level for different developmental stages and under various environmental conditions. To date, the genome sequence (The Arabidopsis Genome Initiative, 2000), the epigenomic status (Cokus et al., 2008; Chodavarapu et al., 2010; Roudier et al., 2011; Pascuzzi et al., 2014; Sequeira-Mendes et al., 2014), three-dimensional chromosome structures in the nucleus based on FISH and Hi-C analyses (Pecinka et al., 2004; Schubert et al., 2012, 2014; Feng et al., 2014; Grob et al., 2014; Wang et al., 2015), and extensive transcriptome data from DNA microarrays have been obtained from *Arabidopsis* and are available from public databases such as Gene Expression Omnibus (GEO) and Array Express. The exhaustive analysis of these transcriptome data sets has provided insights into intermolecular relationships in stress responses (Maruyama-Nakashita et al., 2005; Kilian et al., 2007), the responses to hormone treatment (Nemhauser et al., 2006; Goda et al., 2008), and amino acid metabolism (Less and Galili, 2008; Wittenberg et al., 2012) in *Arabidopsis*.

Recently, the phenotypic plasticity of plants, from short-term plasticity such as changes in leaf angle, stomatal aperture, and photosynthetic rate to long-term plasticity such as developmental and cross-generational plasticity against short and long-term environmental variations, has been extensively investigated (Sultan, 2000; Valladares et al., 2007; Matesanz et al., 2010). Phenotypic variance has often been observed in plants with the same genome sets and grown under the same environmental conditions. Analyses of the gene expression noise that may be the origin of this phenotypic variation have been undertaken in rice and *Arabidopsis* (Nagano et al., 2012; Shen et al., 2012). In rice, these analyses indicated a significant relationship between gene expression noise and the functions of the genes (Nagano et al., 2012). In this paper, transcriptome data sets from *Arabidopsis* were used to search for correlations between gene expression noise and short-term plasticity in several tissues, plants of different ages, and under different environmental conditions. Through the analysis of these transcriptome data sets, robust and significant noise-plasticity correlations were identified that were weakly influenced by the properties of the architecture of the promoter sequences, expression levels, and functional groups. These influences differed from those reported in budding yeast and *E. coli*.

**Results**

**Gene expression level variations in different experimental data sets**

The gene expression data sets from *Arabidopsis* were obtained from different tissues from plants of different ages and under variable growth conditions. Expression data for *Arabidopsis* genes were obtained from the Gene Expression Omnibus (GEO: http://www.ncbi.nlm.nih.gov/geo/). In this study, we analyzed well-documented experimental data sets from microarray analyses of 22,746 genes of *Arabidopsis* from the AtGenExpress project (http://www.weigelworld.org/resources/microarray/AtGenExpress). In particular, we analyzed data sets involving control (non-treated wild type *Arabidopsis*, Columbia (Col-0)) data sets, and data sets from treated plants in which the number of time points multiplied by the number of replicates (data samples) of each data set was greater than 8. The names of the data sets, experiments, tissues used, ages (day-old), GEO accession numbers, and numbers of data of expression data sets are given in Table 1.

| Name of data set | Name of experiment | Tissue | Day-old | GEO | # of data (control) | Reference |
|---|---|---|---|---|---|---|
| SS | Stress treatments | Shoot | 18 days | GSE5620 - GSE5628 | Replicates : 2 Time points : 9 | Kilian et al., (2007) |
| SR | Stress treatments | Root | 18 days | GSE5620 - GSE5628 | Repricates : 2 Time points : 9 | Kilian et al., (2007) |
| PS | Pathogen series: Pseudomonas half leaf injection | Leaf | 28 days | GSE5685 | Repricates : 2 Time points : 6 | Dong X, Townsend H, Emmerson Z, Schildknecht B. (2007) |
| ER | Pathogen series: Response to *Erysiphe orontii* infection | Leaf | 31 days | GSE5686 | Replicates : 3 Time points : 8 | Ausubel F, Dewdney J, Townsend H, Emmerson Z, Schildknecht B. (2007) |
| PH | Pathogen series: Response to *Phytophthora infestans* | Leaf | 31 days | GSE5616 | Replicates : 3 Time points : 3 | Scheel D, Brunner F, Westphal L, Townsend H, Emmerson Z, Schildknecht B. (2007) |
| SU | Response to sulfate limitation | Root | 10 days | GSE5688 | Replicates : 2 Time points : 6 | Maruyama et al., (2005) |

Table 1. Names of data sets, names of experiments, tissues, ages, GEO accession numbers, and the number of replicates and time points of analyzed data sets.

The average expression level and average expression noise of each of 22,746

genes from non-treated (control or mock) conditions were obtained using different values in the different data sets (Supplementary Figure S1). Thus in the following arguments, the noise-plasticity relationships are estimated respectively for each data set.

**Noise-plasticity correlations for *Arabidopsis* gene expression in experimental data sets**

The noise-plasticity relationships of the 22,746 genes for each data set are plotted in Figure 1. Here, the noise ($N$) and plasticity ($P$) of genes for each data set (SS, SR, PS, ER, PH, and SU in Table 1), are named as $N_{SS}$, $N_{SU}$, … $P_{SS}$, … , and $P_{SU}$. The analysis identified significant correlations for all experimental data sets (p-values less than $10^{-4}$). Thus, the noise-plasticity correlations for gene expression were robust and independent of the tissue, ages, growth conditions (see references in Table 1), or treatment.

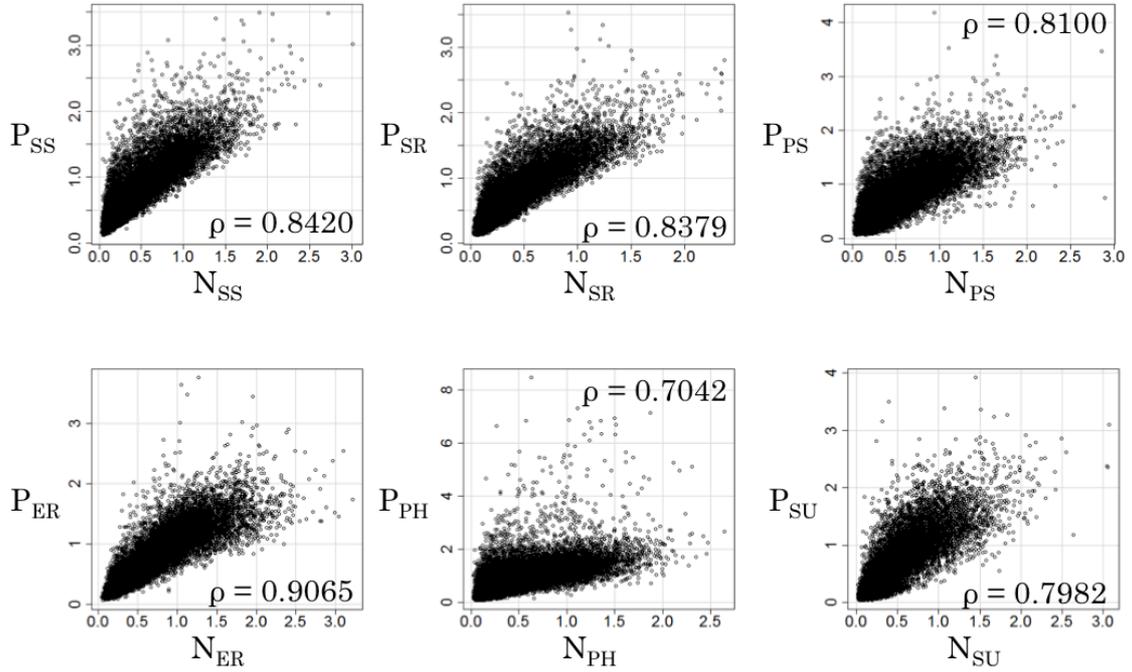

Figure 1: Noise-plasticity scatter plots for 22,746 *Arabidopsis* genes from the SS, SR, PS, ER, PH, and SU data sets (Table 1). Here, the noise ($N$) and plasticity ($P$) of genes for each data set are named as $N_{SS}$, $N_{SU}$, … $P_{SS}$, … , and $P_{SU}$. For all correlation coefficients ($\rho$), a p-value of less than $10^{-4}$ was obtained.

**Promoter architecture dependency of noise-plasticity correlations**

The noise-plasticity correlations of gene groups with and without TATA promoters were estimated for each experimental data set (Table 2) from the noise-plasticity scatter plots (Figure 2 and Supplementary Figure S2). The analysis indicated that significant noise-plasticity correlations were present in both gene groups although the correlation coefficient for the gene group with a TATA box was a little stronger.

| Gene class | | Data Sets | | | | | | # of genes |
|---|---|---|---|---|---|---|---|---|
| | | SS | SR | PS | ER | PH | SU | |
| Motif Search | TATA | 0.801162547 | 0.812971334 | 0.741665995 | 0.877229303 | 0.638166501 | 0.713678059 | 1707 |
| | non-TATA | 0.70410144 | 0.705280889 | 0.643104428 | 0.805352193 | 0.507874562 | 0.646408494 | 10116 |
| PlantProm | TATA | 0.803969585 | 0.805952095 | 0.72483968 | 0.8693857 | 0.642596505 | 0.713802903 | 3347 |
| | non-TATA | 0.678331305 | 0.683013798 | 0.628029668 | 0.790042142 | 0.473164923 | 0.632156735 | 8476 |

Table 2: Noise-plasticity correlations for gene groups with and without TATA promoters from several data sets (Table 1). For all correlation coefficients, a p-value of less than $10^{-4}$ was obtained.

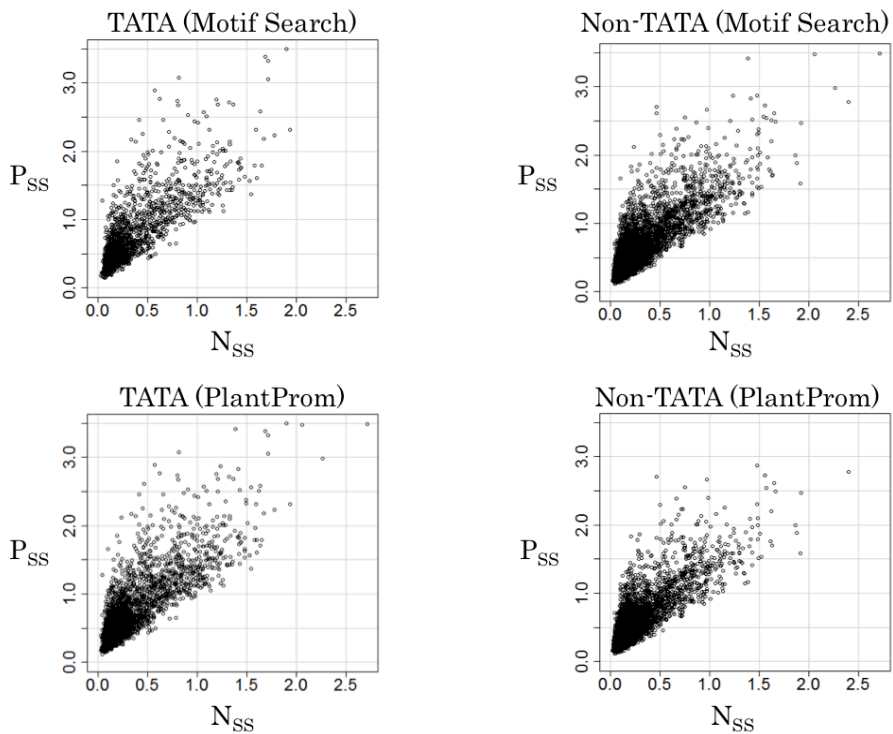

Figure 2: Noise-plasticity scatter plots for the SS data set (Table 1) showing gene

groups with and without TATA promoters.

**Dependency of mean expression level on noise-plasticity correlation**

The noise-plasticity correlations of five gene groups, classified according to average gene expression level under non-treated conditions (Higher, top 20% genes; High, top 20~40% genes; Middle, top 40~60% genes; Low, top 60~80% genes; and Lower, lowest 20% genes), were estimated for each experimental data set (Table 3(a)) from the noise-plasticity scatter plots (Figure 3 and Supplementary Figure S3). The analysis indicated that significant noise-plasticity correlations were present in most of the five gene groups. Moreover, the correlations became slightly weaker with the increase in mean expression levels for the gene group (Table 3(a)). This effect was greater in the gene group without TATA promoters than that with TATA promoters (Table 3(b)(c)).

Notably, average noise levels also correlated negatively with expression levels for each experimental data set (Table 3(d)). This result indicates that gene groups with higher average noise levels exhibited higher noise-plasticity correlations.

(a)

| Gene class | | Data Sets | | | | | | # of genes |
|---|---|---|---|---|---|---|---|---|
| | | SS | SR | PS | ER | PH | SU | |
| Expression levels | Lower | 0.733124838 | 0.783632099 | 0.597850475 | 0.610751316 | 0.525722172 | 0.590437992 | 4549 |
| | Low | 0.751205906 | 0.654080518 | 0.636663247 | 0.776356386 | 0.5567612 | 0.599244736 | 4549 |
| | Middle | 0.563846317 | 0.603224613 | 0.514372871 | 0.677376384 | 0.402774327 | 0.446483382 | 4549 |
| | high | 0.550773716 | 0.583762052 | 0.427086984 | 0.65206279 | 0.353471737 | 0.437797905 | 4549 |
| | higher | 0.602178816 | 0.611146744 | 0.37501576 | 0.643502671 | 0.29192688 | 0.420520585 | 4550 |

(b)

| Gene class | | Data Sets | | | | | | # of genes |
|---|---|---|---|---|---|---|---|---|
| | | SS | SR | PS | ER | PH | SU | |
| Expression levels (TATA) | Lower | 0.760973941 | 0.688469671 | 0.683806494 | 0.81028254 | 0.568021725 | 0.568795428 | 269 |
| | Low | 0.767391945 | 0.683479579 | 0.632377748 | 0.827254237 | 0.608930353 | 0.539644141 | 254 |
| | Middle | 0.684286234 | 0.660382937 | 0.689429698 | 0.850152684 | 0.599622731 | 0.502088091 | 305 |
| | high | 0.716276983 | 0.69344424 | 0.658948087 | 0.843240439 | 0.614008526 | 0.480777249 | 295 |
| | higher | 0.787183283 | 0.742093661 | 0.647655171 | 0.832084328 | 0.528034686 | 0.313427815 | 584 |

(c)

| Gene class | | Data Sets | | | | | | # of genes |
|---|---|---|---|---|---|---|---|---|
| | | SS | SR | PS | ER | PH | SU | |
| Expression levels (non-TATA) | Lower | 0.837761235 | 0.746248624 | 0.740161065 | 0.874956806 | 0.565182476 | 0.575050951 | 609 |
| | Low | 0.728100781 | 0.67583862 | 0.679686766 | 0.871241636 | 0.499103193 | 0.613526895 | 1228 |
| | Middle | 0.567621274 | 0.54934547 | 0.569903096 | 0.789535519 | 0.380529313 | 0.430334001 | 2185 |
| | high | 0.599047866 | 0.546126369 | 0.506526437 | 0.71558567 | 0.379211437 | 0.416208194 | 2828 |
| | higher | 0.646805935 | 0.549960964 | 0.496327819 | 0.706405794 | 0.416916505 | 0.424672755 | 3266 |

(d) Noise

| Gene class | | Data Sets | | | | | | # of genes |
|---|---|---|---|---|---|---|---|---|
| | | SS | SR | PS | ER | PH | SU | |
| Expression levels | Lower | 0.833143589 | 0.775106015 | 0.871663739 | 1.183791487 | 0.906237011 | 0.807047097 | 4549 |
| | Low | 0.421476797 | 0.298163723 | 0.546898518 | 0.644063817 | 0.585149829 | 0.340627556 | 4549 |
| | Middle | 0.205919853 | 0.181029325 | 0.265538034 | 0.321443824 | 0.270158729 | 0.203542309 | 4549 |
| | high | 0.14762109 | 0.14011307 | 0.186279601 | 0.238927916 | 0.183552864 | 0.170186527 | 4549 |
| | higher | 0.118904418 | 0.121164766 | 0.163101715 | 0.219011572 | 0.150074919 | 0.148419355 | 4550 |

Table 3: (a) Noise-plasticity correlations among five gene groups classified by average gene expression levels from several data sets (Table 1). (b)(c) Noise-plasticity correlations of gene groups with and without TATA promoters belong to gene groups with different expression levels. For all correlation coefficients, a p-value of less than $10^{-4}$ was obtained. (b) Noise strength of five gene groups classified by average gene expression levels from several data sets.

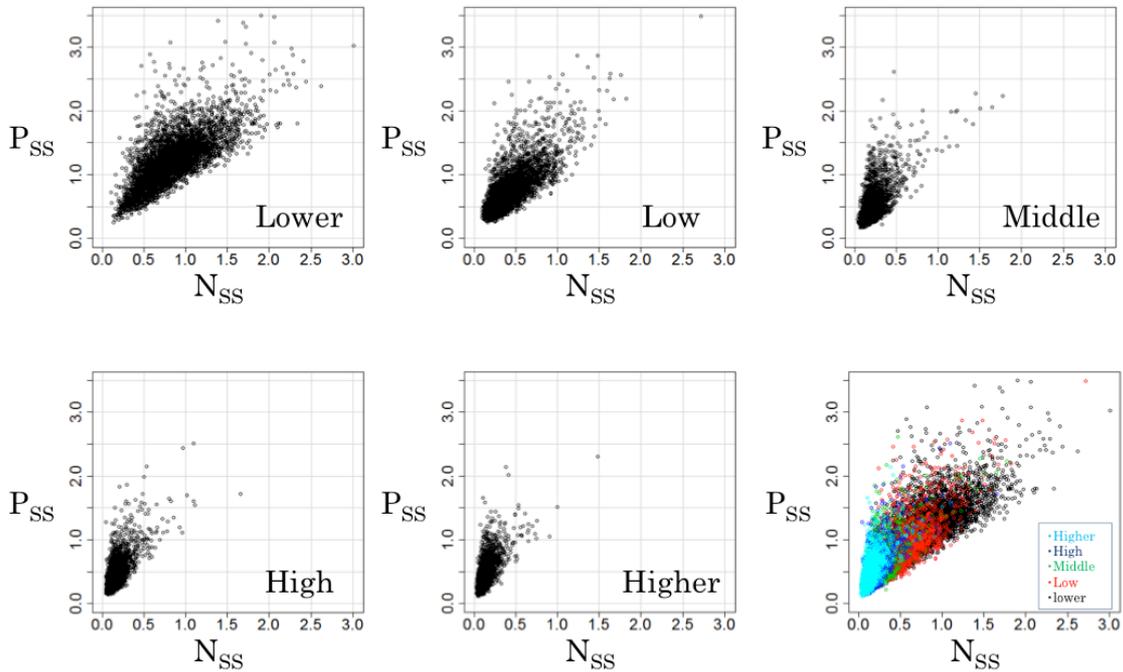

Figure 3: Noise-plasticity scatter plots for the SS data set (Table 1) for the five gene groups classified by average gene expression level.

**Dependency of gene function on noise-plasticity correlations**

The noise-plasticity correlations of gene groups classified as "Essential" genes (Meinke et al., 2008) were estimated for each experimental data set (Table 4(a)). Strong noise-plasticity correlations were obtained from most of the data sets except the PH data set. Next, the 22,746 genes were classified into the Gene Ontology (GO) Slim terms from TAIR; the noise-plasticity correlations were then estimated for each group.

Significant noise-plasticity correlations were found regardless of the nature of the gene group, and high correlation coefficients were always obtained for groups containing more than 1000 genes (Table 4(b) and Supplementary Table S1). However, low correlation coefficients were obtained in some small gene groups, e.g., the PH data set classified as "generation of precursor metabolites and energy."

| (a) | Data Sets | | | | | | |
|---|---|---|---|---|---|---|---|
| Gene class | SS | SR | PS | ER | PH | SU | # of genes |
| Essential genes | 0.594221533 | 0.620498722 | 0.611344186 | 0.779395159 | 0.310282866 | 0.561288897 | 481 |

| (b) | | Data Sets | | | | | | |
|---|---|---|---|---|---|---|---|---|
| Gene class | | SS | SR | PS | ER | PH | SU | # of genes |
| GO Slim biological process | multicellular organismal development | 0.777901784 | 0.77590979 | 0.76517542 | 0.879061975 | 0.618875647 | 0.721621913 | 3408 |
| | transport | 0.806768388 | 0.779186949 | 0.719216124 | 0.861581653 | 0.615249087 | 0.68722564 | 3276 |
| | signal transduction | 0.820602108 | 0.792641264 | 0.692514902 | 0.891325264 | 0.693661654 | 0.761492046 | 1831 |
| | cellular component organization | 0.765876131 | 0.752584687 | 0.722483496 | 0.856916337 | 0.57692963 | 0.707008042 | 3456 |
| | cellular process | 0.798034579 | 0.782971878 | 0.751303878 | 0.877148094 | 0.634387429 | 0.732883121 | 12297 |
| | DNA metabolic process | 0.782529383 | 0.666095033 | 0.844019501 | 0.913947662 | 0.73702052 | 0.612254361 | 804 |
| | RNA localization | 0.736251896 | 0.459853141 | 0.749062176 | 0.689480405 | 0.407095538 | 0.582648256 | 67 |
| | protein metabolic process | 0.789746546 | 0.75550503 | 0.7378335 | 0.862661771 | 0.626479244 | 0.708425593 | 4183 |
| | generation of precursor metabolites and energy | 0.637920223 | 0.793310459 | 0.479428222 | 0.687023532 | 0.221118249 | 0.753399527 | 571 |
| | transcription, DNA-dependent | 0.808018159 | 0.796249517 | 0.780831718 | 0.900260213 | 0.703566562 | 0.763686382 | 2228 |
| | metabolic process | 0.80529708 | 0.78527943 | 0.757569657 | 0.879157491 | 0.640619377 | 0.734568262 | 11591 |
| | response to abiotic stimulus | 0.765058976 | 0.760153724 | 0.67062848 | 0.844872619 | 0.589955057 | 0.679993856 | 2682 |
| | response to biotic stimulus | 0.832086965 | 0.764543076 | 0.599308809 | 0.814099573 | 0.604496842 | 0.707391195 | 1374 |
| | response to stress | 0.821338172 | 0.778983491 | 0.698520265 | 0.867255788 | 0.643517632 | 0.710583396 | 3857 |

Table 4: (a) Noise-plasticity correlations among gene groups classified as essential

genes, and (b) those among gene groups classified by Gene Ontology (GO) terms (Biological processes) from several data sets (Table 1). For all correlation coefficients, p-values of less than $10^{-4}$ were obtained.

Genes exhibiting high responses to plant hormones [indole acetic acid (IAA), zeatin, gibberellic acid (GA), 1-aminocyclopropane-1-caboxylic acid (ACC), brassinolide (BL), abscisic acid (ABA) and methyl jasmonate (mJA)] were grouped and their noise-plasticity correlations estimated in each data set. Strong noise-plasticity correlations were obtained for all gene groups (Table 5).

| Data Sets | | | | | | | |
|---|---|---|---|---|---|---|---|
| Gene class | | SS | SR | PS | ER | PH | SU | # of genes |
| Hormone response genes | IAA | 0.895791776 | 0.897672319 | 0.802838469 | 0.893207293 | 0.723767073 | 0.862105274 | 4550 |
| | Zeatin | 0.825813936 | 0.848923607 | 0.677984814 | 0.793822528 | 0.566258366 | 0.778086978 | 4550 |
| | GA | 0.819367447 | 0.844356479 | 0.6676091 | 0.785805148 | 0.564767919 | 0.76607735 | 4550 |
| | ABA | 0.781056696 | 0.827645964 | 0.700394021 | 0.824830726 | 0.595547395 | 0.80895369 | 4550 |
| | MJ | 0.808847447 | 0.842218737 | 0.696383404 | 0.818964635 | 0.565759064 | 0.799685123 | 4550 |
| | ACC | 0.82858931 | 0.848031519 | 0.672945983 | 0.800435685 | 0.574560917 | 0.773563269 | 4550 |
| | BL | 0.816362997 | 0.849880114 | 0.684714359 | 0.8016745 | 0.577559444 | 0.771162878 | 4550 |

Table 5: Noise-plasticity correlations of the gene groups showing plant hormone responsiveness for several data sets (Table 1). For all correlation coefficients, p-values of less than $10^{-4}$ were obtained.

**Relationship between circadian rhythms and noise-plasticity correlations**

The 22,746 genes were assigned to five gene groups on the basis of the correlation between their expression level and circadian rhythms (Higher, top 20% genes; High, top 20~40% genes; Middle, top 40~60% genes; Low, top 60~80% genes; and Lower, lowest 20% genes). Correlation coefficients were then estimated for each experimental data set (Table 6) from the noise-plasticity scatter plots (Figure 4 and Supplementary Figure S4). This analysis indicated that significant noise-plasticity correlations were present in all gene groups. However, the gene groups with a stronger correlation between gene expression level and circadian rhythm tended to show weaker noise-plasticity correlation than other groups.

| Gene class | | Data Sets | | | | | | # of genes |
|---|---|---|---|---|---|---|---|---|
| | | SS | SR | PS | ER | PH | SU | |
| Correlations | Lower | 0.920045389 | 0.891384101 | 0.860561377 | 0.937665389 | 0.795718454 | 0.893298791 | 4545 |
| with | Low | 0.903607033 | 0.872385705 | 0.836301432 | 0.928875391 | 0.750849369 | 0.850201092 | 4544 |
| circadian | Middle | 0.852512795 | 0.846553647 | 0.809667114 | 0.922343383 | 0.682079987 | 0.804015191 | 4544 |
| rhythm | high | 0.784579079 | 0.8002751 | 0.773660065 | 0.899409403 | 0.568273608 | 0.721750279 | 4544 |
| | higher | 0.684126264 | 0.70732506 | 0.727442656 | 0.776911284 | 0.474371623 | 0.619838995 | 4545 |

Table 6: Noise-plasticity correlations among five gene groups classified by the correlations between their expression levels and circadian rhythm for several data sets (Table 1). For all correlation coefficients, p-values of less than $10^{-4}$ were obtained.

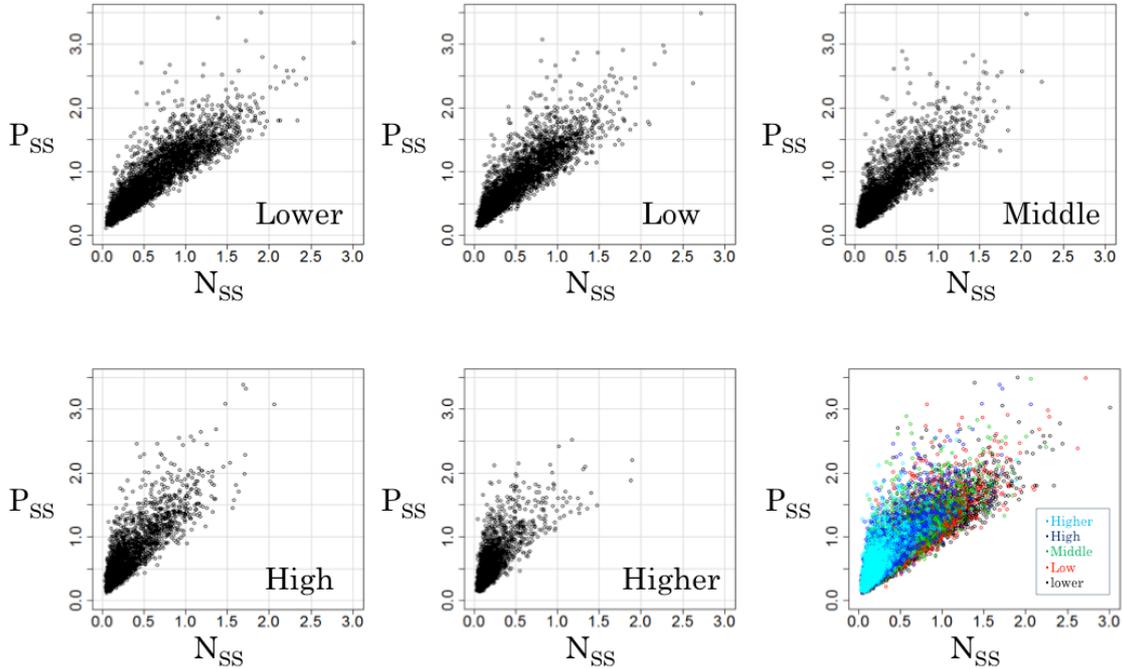

Figure 4: Noise-plasticity scatter plots of the five gene groups classified by the correlation between their expression levels and circadian rhythms for the SS data set (Table 1).

## Discussion

Recent studies in budding yeast identified several gene groups that exhibit strong noise-plasticity correlations and others that do not show such correlations. For example, gene groups in which each gene has a promoter with a TATA box show strong noise-plasticity correlation for expression levels; by contrast, those without a TATA box do not show this correlation (Lehner, 2010). In the present study on *Arabidopsis,* the noise-plasticity correlations of groups containing genes with TATA promoters were similarly larger than those without TATA promoters. However, the difference in the strength of the noise-plasticity correlations in *Arabidopsis* was much smaller than found in budding yeast (Table 7).

In *E. coli*, only gene groups with higher mean expression levels exhibit significant correlations between expression noise and plasticity (Singh, 2013). In contrast, in *Arabidopsis*, strong noise-plasticity correlations were obtained for most gene groups, and gene groups with low mean expression levels exhibited particularly strong correlations (Table 3(a), Table 7). Moreover, the negative correlations between the noise-plasticity correlation coefficients and mean expression levels were greater for the gene group without TATA promoters in *Arabidopsis* (Table 3(b)(c)). On the other hand, little correlation between the noise-plasticity correlation coefficients and mean expression levels has been reported for both gene groups with and without TATA promoters in budding yeast (Lehner, 2010) (Table 7). Thus, the influence of mean expression level on noise-plasticity relationships might differ greatly from that in unicellular organisms.

Furthermore, in budding yeast and *E. coli*, gene groups with essential functions in cell growth and those with dosage-sensitive functions do not show strong noise-plasticity correlations (Lehner, 2010; Singh, 2013). Indeed, the noise-plasticity correlations for essential gene groups were weaker than those for all genes in *Arabidopsis*. However, in *Arabidopsis*, stronger noise-plasticity correlations were found for essential gene groups compared to budding yeast and *E. coli* (Table 4, Table 7), and were also found for most gene groups classified by GO Slim terms (Table 7, and Supplementary Tables S1). Moreover, strong noise-plasticity correlations were obtained for several gene groups exhibiting sensitivity to plant hormones (both growth hormones and stress hormones). Thus, strong noise-plasticity correlations are expected in most gene groups independent of their functional roles in *Arabidopsis*.

**(a)**

| Gene class | Yeast (Lehner 2010) | | E. coli (Singh 2013) | | Arabidopsis (SR) | |
|---|---|---|---|---|---|---|
| All genes | | 0.3 | | 0.14 | | 0.84 |
| TATA genes | | 0.62 | ************** | | | 0.8 |
| non-TATA genes | | 0.16 | ************** | | | 0.71 |
| Lower (expression level) | TATA | 0.58 | | | All | 0.78 |
| | nonTATA | 0.19 | All | 0.06 | TATA | 0.69 |
| | | | | | non-TATA | 0.75 |
| Low (expression level) | TATA | 0.70 | | | All | 0.65 |
| | nonTATA | 0.24 | All | 0.08 | TATA | 0.68 |
| | | | | | non-TATA | 0.66 |
| Middle (expression level) | TATA | 0.68 | | | All | 0.60 |
| | nonTATA | 0.20 | All | 0.16 | TATA | 0.66 |
| | | | | | non-TATA | 0.55 |
| High (expression level) | TATA | 0.70 | | | All | 0.58 |
| | nonTATA | 0.28 | All | 0.18 | TATA | 0.69 |
| | | | | | non-TATA | 0.55 |
| Higher (expression level) | TATA | 0.61 | | | All | 0.61 |
| | nonTATA | 0.25 | All | 0.23 | TATA | 0.74 |
| | | | | | non-TATA | 0.55 |
| Essential genes | | 0.15 | | -0.01 | | 0.62 |

**(b)**

| Gene class (GO slim) | Yeast (Lehner 2010) | | Arabidopsis (SR) | |
|---|---|---|---|---|
| molecular_function | | | All | 0.82 |
| | TATA | 0.65 | TATA | 0.81 |
| | nonTATA | 0.18 | non-TATA | 0.71 |
| structural molecule activity | | | All | 0.64 |
| | TATA | 0.57 | TATA | 0.49 |
| | nonTATA | -0.37 | non-TATA | 0.46 |
| transporter activity | | | All | 0.81 |
| | TATA | 0.46 | TATA | 0.74 |
| | nonTATA | 0.37 | non-TATA | 0.74 |
| protein binding | | | All | 0.75 |
| | TATA | 0.54 | TATA | 0.84 |
| | nonTATA | 0.1 | non-TATA | 0.68 |
| cell wall | | | All | 0.82 |
| | TATA | 0.20 | TATA | 0.84 |
| | nonTATA | 0.14 | non-TATA | 0.73 |
| nucleus | | | All | 0.81 |
| | TATA | 0.57 | TATA | 0.80 |
| | nonTATA | 0.14 | non-TATA | 0.70 |
| cytoplasm | | | All | 0.78 |
| | TATA | 0.57 | TATA | 0.82 |
| | nonTATA | 0.15 | non-TATA | 0.67 |
| mitochondrion | | | All | 0.82 |
| | TATA | 0.48 | TATA | 0.84 |
| | nonTATA | 0.32 | non-TATA | 0.65 |
| ribosome | | | All | 0.60 |
| | TATA | 0.30 | TATA | 0.61 |
| | nonTATA | -0.29 | non-TATA | 0.45 |

| Gene class (GO slim) | Yeast (Lehner 2010) | | Arabidopsis (SR) | |
|---|---|---|---|---|
| plasma membrane | | | All | 0.77 |
| | TATA | 0.51 | TATA | 0.79 |
| | nonTATA | 0.50 | non-TATA | 0.70 |
| generation of precursor metabolites and | | | All | 0.79 |
| | TATA | 0.24 | TATA | 0.89 |
| | nonTATA | 0.50 | non-TATA | 0.76 |
| transcription | | | All | 0.80 |
| | TATA | 0.76 | TATA | 0.78 |
| | nonTATA | 0.07 | non-TATA | 0.70 |
| transport | | | All | 0.78 |
| | TATA | 0.59 | TATA | 0.79 |
| | nonTATA | 0.25 | non-TATA | 0.71 |
| response to stress | | | All | 0.78 |
| | TATA | 0.68 | TATA | 0.82 |
| | nonTATA | 0.16 | non-TATA | 0.72 |
| signal transduction | | | All | 0.79 |
| | TATA | 0.62 | TATA | 0.86 |
| | nonTATA | 0.31 | non-TATA | 0.72 |
| biological_process | | | All | 0.82 |
| | TATA | 0.67 | TATA | 0.81 |
| | nonTATA | 0.28 | non-TATA | 0.71 |
| membrane | | | All | 0.77 |
| | TATA | 0.57 | TATA | 0.78 |
| | nonTATA | 0.29 | non-TATA | 0.69 |
| transferase activity | | | All | 0.76 |
| | TATA | 0.56 | TATA | 0.79 |
| | nonTATA | 0.12 | non-TATA | 0.65 |

Table 7: (a) Comparison of noise-plasticity correlation strength for (a) all genes,

TATA and, non-TATA gene groups, gene groups with different expression levels among budding yeast, *E. coli*, and *Arabidopsis*, and (b) for common GO slim terms between budding yeast and *Arabidopsis*,. The results from data set SR (Table 1) are used as typical results of *Arabidopsis*.

However, the strength of the correlation between gene expression level and circadian rhythm did influence noise-plasticity relationships in *Arabidopsis*. Gene groups with a high correlation with the circadian rhythm exhibited lower noise-plasticity correlations than other gene groups. Thus, while the functional roles of the genes did not influence noise-plasticity relationships, the correlation with circadian rhythm did appear to be essential to this relationship in *Arabidopsis*.

There were some exceptions to general robustness of noise-plasticity correlations.

For example, noise-plasticity correlations for infection with *Phytophthora infestans*

(from PH data sets) tended to be weaker compared to those for the other treatment groups (Table 2, Table 3). In the PH data set, gene groups that had high mean expression levels and that were classified in the GO term "generation of precursor metabolites and energy" exhibited comparatively weak noise-plasticity correlations for the pathogen (Figure 5). This finding indicates that the strength of the noise-plasticity correlations depends on the properties of the treatment.

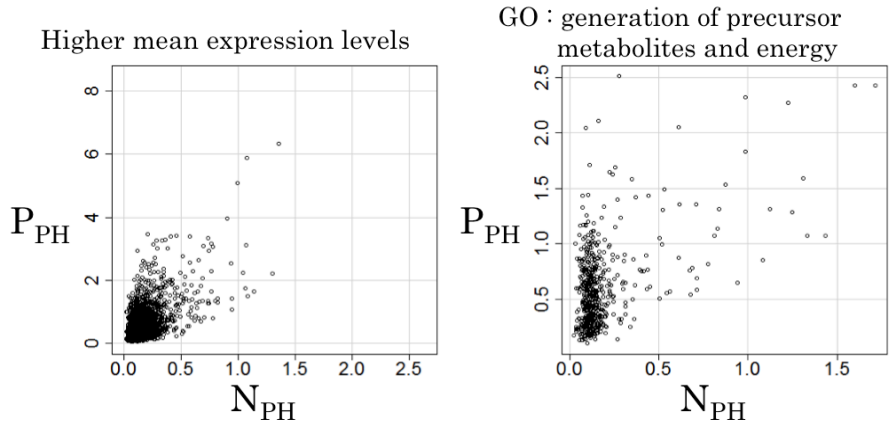

Figure 5: Noise-plasticity relationship of the gene group with high mean expression

levels and classified in the GO term "generation of precursor metabolites and energy."

This group exhibited weak noise-plasticity correlations for *Phytophthora infestans*.

(PH, Table 1)

The results described in the present report suggest that the characteristics of noise-plasticity correlations in *Arabidopsis* differ from those in budding yeast and *E.

coli*. In *Arabidopsis*, noise-plasticity correlations are essentially robust at different ages and in different tissues, and are only weakly influenced by promoter architecture, mean expression level, and circadian rhythm.

Multicellular organisms exhibit different circadian rhythms, cell differentiation patterns, and cell-cell interactions from primitive unicellular organisms such as yeast and *E. coli*. Moreover, multicellular organisms also show more complex higher order chromosome structures. For example, *Arabidopsis* chromosomes are considerably larger than those of the budding yeast: the genome size of the smallest chromosome of *Arabidopsis* (chromosome 4) is ~18 Mb (total genome size ~120 Mb) while that of the largest chromosome (chromosome 4) of budding yeast is 1.5 Mb (total genome size ~12 Mb). Thus, the *Arabidopsis* chromosomes are expected to contain more complex hierarchical structures. FISH and Hi-C analyses have identified topologically associated domain-like structures in *Arabidopsis* that are absent in budding yeast (Pecinka et al., 2004; Duan et al., 2010; Schubert et al., 2012, 2014; Feng et al., 2014; Grob et al., 2014; Wang et al., 2015).

Thus, we conjecture that gene regulation in *Arabidopsis* is more influenced by circadian rhythm-mediated global temporal regulation and by global structural complexities and transitions in chromosomes than by the local characteristics of the loci as in yeast and *E. coli*. This argument also suggests that noise-plasticity relationships differ between unicellular and multicellular organisms. This prediction needs to be further tested using gene expression data from a wider range of organisms.

**Materials and Methods**

**Data sources for gene classification**

Data on promoter architectures, i.e., whether or not they include TATA boxes, were obtained from The *Arabidopsis* Gene Regulatory Information Server (AGRIS: http://arabidopsis.med.ohio-state.edu). Here, two types of classification for promoter architecture are provided, namely, Motif Search and PlantProm. In order to classify each gene, the complete Gene Ontology (GO) biological classification list and GO Slim classification list were obtained from TAIR (http://www.arabidopsis.org). Data on "Essential" genes were obtained from SeedGenes Project (Meinke et al., 2008, http://www.seedgenes.org/GeneList).

**Definitions and estimations of gene expression noise and plasticity**

The growth conditions and ages differed among the experimental data sets. Thus,

for each data set, the noise-plasticity relationship was measured individually.

Recent gene expression analyses indicated that from *E. coli* to mammalian cells, the fluctuations in expression level for each gene tend to follow a log-normal distribution (Sato et al., 2003; Furusawa et al., 2005; Chang et al., 2008; Konishi et al., 2008). In this case, the log of the gene expression level is generally considered suitable for various statistical analyses since its distribution function is essentially Gaussian. Thus, expression noise and plasticity for each gene are defined and estimated using the deviations among $\log_2$[microarray signal intensity] as described in an earlier study (Sato et al., 2003). These estimations reveal that the values for noise and plasticity have no dimensions and are not influenced by gene specific characteristic scales such as average expression levels.

The expression noise for each gene is defined as the stochastic variation among samples of $\log_2$ intensities under non-treated conditions (control data, "Mock" condition data, etc.). For two or more sets of data for microarray signal intensities obtained under the same conditions at the same time point $t$, $C_i(t)$ ($i = 1, 2,$ or 3 indicates sample index), the expression noise for each gene is estimated by the time average of standard deviations among the sample data, $Std\{\ \}_i$, as

$N = \langle Std\{log_2 C_i(t)\}_i \rangle_t$ where $\langle\ \rangle_t$ indicates the average over $t$ (Fig. 6(a)). In

addition, the average expression levels of each gene under non-treated conditions are defined as $A = \langle\langle log_2 C_i(t) \rangle_i\rangle_t$ where $\langle\ \rangle_i$ indicates the average over $i$

The expression plasticity (P) for each gene is defined by the absolute variation

between sample averages of $log_2$ intensities under stress and non-treated (control or

Mock) conditions, $\langle log_2 T_i(t) \rangle_i$ and $\langle log_2 C_i(t) \rangle_i$, where $T_i(t)$ indicates the $i$-th

sample of the microarray signal intensity under a treated condition at time t. The

plasticity is estimated as the time average of absolute deviations between the sample

average of $log_2$ intensities of the treated and non-treated conditions measured as

$P = \langle |\langle log_2 T_i(t) \rangle_i - \langle log_2 C_i(t) \rangle_i| \rangle_t$ where $\langle\ \rangle_i$ indicates the average over $i$ (Fig.

6(b)).

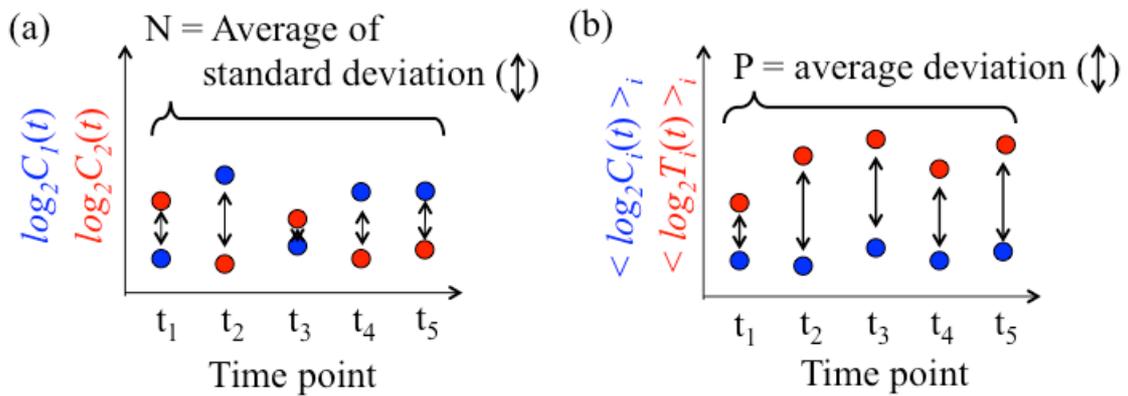

Figure 6: Illustrations of the definitions of (a) noise N, and (b) plasticity P.

The noise ($N$) and plasticity ($P$) of genes for each data set are named as $N_{ds}$ and $P_{ds}$, such as $N_{SS}$, $N_{SU}$, ... $P_{SS}$ and $P_{SU}$. Note that the data sets SS and SR involve the data of 8 stress treatments (cold, osmotic shock, salt, drought, genomic, UV, wound and heat stresses). Thus, $P_{SS}$ and $P_{SR}$ are estimated as the average of $P$ among these treatments.

**Selection of hormone-response genes**

A microarray analysis data set on plant hormone responses is available in AtGenExpress and was used here to identify plant hormone-related gene groups. The same 22,746 genes as in Table 1 were used here; two samples of three time points data were obtained for control and each hormone applied (IAA, zeatin, GA, ACC, BL, ABA and mJA) (Goda et al., 2008).

The hormone response (R) for each gene is defined by the absolute variation between sample averages of $\log_2$ intensities under hormone-treated and non-treated conditions, $\langle log_2 A_i(t)\rangle_i$ and $\langle log_2 C_i(t)\rangle_i$, where $A_i(t)$ indicates the $i$-th sample of the microarray signal intensity under a treatment condition at time t. The hormone response is estimated as the time average of absolute deviations between the sample average of $log_2$ intensities in the treated and non-treated conditions measured as $R = \langle|\langle log_2 A_i(t)\rangle_i - \langle log_2 C_i(t)\rangle_i|\rangle_t$, where $\langle \ \ \rangle_i$ and $\langle \ \ \rangle_t$ indicate the average

over $i$ and time point $t$. For each hormone, genes with the top 20% for $R$ were selected as hormone-response genes (more than 90% of selected genes for each hormone treatment hold $R > 1$).

**Definitions of correlation between gene expression levels and circadian rhythm**

For each data set, the correlation between each gene expression level and

circadian rhythm was defined as the maximum value among the absolute values of

Pearson correlation coefficients between the expression levels of the gene and one of the clock-associated genes in the control conditions. Here, *CCA1*, *LHY1*, *PRR9*, *PRR7*,

*PRR5*, *TOC1*, *LUX*, *ELF4*, and *ELF3* were chosen as clock-associated genes since they are known to constitute the core regulatory network of the circadian rhythm in *Arabidopsis* (Nakamichi, 2011).

## Correlation estimations

The correlations among several values were estimated using Spearman correlation coefficients. The statistical analyses were performed using R (http://www.r-project.org).

## Acknowledgements

The authors are grateful to A. Sakamoto, H. Kudoh, S. Watanabe, and H. Nishimori for fruitful discussions.

Funding:

**Funder:** Platform Project for Supporting in Drug Discovery and Life Science Research (Platform for Dynamic Approaches to Living System) from the Ministry of Education, Culture, Sports, Science (MEXT) and Japan Agency for Medical Research and development (AMED).

**Grant reference number:**

**Author:** Akinori Awazu

**Funder:** Grant-in-Aid for Scientific Research on Innovative Areas "Integrated Analysis of Strategies for Plant Survival and Growth in Response to Global Environmental Changes" of MEXT of Japan

**Grant reference number:** 25119718

**Author:** Akinori Awazu

**Funder:** Grant-in-Aid for Scientific Research on Innovative Areas "Initiative for High-Dimensional Data-Driven Science through Deepening of Sparse Modeling" of MEXT of Japan

**Grant reference number:** 26120525

**Author:** Akinori Awazu

**Funder:** PRESTO of Japan Science and Technology Agency

**Grant reference number:**

**Author:** Atsushi J. Nagano

**Author Contributions**

A.A. and A.J.N. conceived the project. K.H. and A.A. performed the data analysis, and

A.A. and A.J.N. wrote the manuscript.